\documentclass[reprint,aps,prl,amsmath,amssymb,superscriptaddress,floatfix]{revtex4-2}

\usepackage{amstext}
\usepackage{graphicx}
\usepackage{color}
\usepackage{xspace}
\usepackage{dcolumn}
\usepackage{bm}
\usepackage[dvipsnames]{xcolor}
\usepackage{soul}
\usepackage{gensymb}
\usepackage[tight]{units} 
\usepackage{textcomp}

\begin{document}

\title{Creep-enhanced vortex pinning revealed through nonmonotonic relaxation of the Campbell length}

\author{Sunil~Ghimire}
\affiliation{Ames National Laboratory, Ames, Iowa 50011, USA}
\affiliation{Department of Physics \& Astronomy, Iowa State University, Ames, Iowa 50011, USA}

\author{Filippo Gaggioli}
\affiliation{Department of Physics, Massachusetts Institute of Technology, Cambridge, MA-02139,USA}
\affiliation{Institut f\"ur Theoretische Physik, ETH Z\"urich, CH-8093 Z\"urich, Switzerland}

\author{Kamal~R.~Joshi}
\affiliation{Ames National Laboratory, Ames, Iowa 50011, USA}

\author{Marcin~Ko\'{n}czykowski}
\affiliation{Laboratoire des Solides Irradi\'{e}s, CEA/DRF/lRAMIS, \'{E}cole Polytechnique, CNRS, Institut Polytechnique de Paris, F-91128 Palaiseau, France}

\author{Romain~Grasset}
\affiliation{Laboratoire des Solides Irradi\'{e}s, CEA/DRF/lRAMIS, \'{E}cole Polytechnique, CNRS, Institut Polytechnique de Paris, F-91128 Palaiseau, France}

\author{Elizabeth~H.~Krenkel}
\affiliation{Ames National Laboratory, Ames, Iowa 50011, USA}
\affiliation{Department of Physics \& Astronomy, Iowa State University, Ames, Iowa 50011, USA}

\author{Amlan~Datta}
\affiliation{Ames National Laboratory, Ames, Iowa 50011, USA}
\affiliation{Department of Physics \& Astronomy, Iowa State University, Ames, Iowa 50011, USA}

\author{Makariy~A.~Tanatar}
\affiliation{Ames National Laboratory, Ames, Iowa 50011, USA}
\affiliation{Department of Physics \& Astronomy, Iowa State University, Ames, Iowa 50011, USA}

\author{Shuzhang~Chen}
\affiliation{Condensed Matter Physics and Materials Science Department, Brookhaven National Laboratory, Upton, New York 11973, USA}
\affiliation{Department of Physics and Astronomy, Stony Brook University, Stony Brook, New York 11794-3800, USA}

\author{Cedomir~Petrovic}
\affiliation{Condensed Matter Physics and Materials Science Department, Brookhaven National Laboratory, Upton, New York 11973, USA}
\affiliation{Department of Physics and Astronomy, Stony Brook University, Stony Brook, New York 11794-3800, USA}

\author{Vadim~B.~Geshkenbein}
\affiliation{Institut f\"ur Theoretische Physik, ETH Z\"urich, CH-8093 Z\"urich, Switzerland}

\author{Ruslan~Prozorov}
\email[Corresponding author: ]{prozorov@ameslab.gov}
\affiliation{Ames National Laboratory, Ames, Iowa 50011, USA}
\affiliation{Department of Physics \& Astronomy, Iowa State University, Ames, Iowa 50011, USA}

\date{31 May 2024}

\begin{abstract}

We study the effects of flux creep on the linear AC response of the vortex lattice in single crystals Ca$_3$Ir$_4$Sn$_{13}$ by measuring the Campbell penetration depth, $\lambda_{\rm \scriptscriptstyle C}(T,H,t)$.
Thermal fluctuations release vortices from shallow pinning sites, only for them to become re-trapped by deeper potential wells, causing an initial increase of the effective Labusch parameter, which is proportional to the pinning well curvature.  This effect cannot be detected in conventional magnetic relaxation measurements but is revealed by our observation of a nonmonotonic time evolution of $\lambda_{\rm \scriptscriptstyle C}(T,H,t)$, which directly probes the average curvature of the occupied pinning centers. The time evolution of $\lambda_{\rm \scriptscriptstyle C}(T,H,t)$ was measured at different temperatures in samples with different densities of pinning centers produced by electron irradiation. The curves can be collapsed together when plotted on a logarithmic time scale $t \to T\ln{(t/t_0)}$ confirming that the time evolution is driven by flux creep. The $\lambda_{\rm \scriptscriptstyle C}(T,H,t)$ is hysteretic with a noticeable nonmonotonic relaxation in the presence of a vortex density gradient (after zero-field cooling), but is monotonic after field cooling, where the vortex density is uniform. This result quantitatively corroborates the novel picture of vortex creep based on the strong pinning theory.
\end{abstract}

\maketitle

\section{Introduction}

The interaction of Abrikosov vortices \cite{Abrikosov1957} with defects, called vortex pinning, determines the magnetic properties of type-II superconductors and the amount of supercurrent they can carry without dissipation \cite{CampbellEvetts2006,Blatter1994}. The usually measured DC magnetization or conventional amplitude-domain AC susceptibility reflect the macroscopic response of the entire sample with vortices moving over large distances, and it is difficult to link them directly to the microscopic vortex pinning and creep mechanisms. This is where the vortex lattice reaction to a very small oscillating external magnetic field that does not force vortices out of their pinning wells becomes very useful. In this regime, the AC magnetic field perturbation is exponentially damped by the vortex lattice on a characteristic length scale called the Campbell penetration depth, $\lambda_{\rm \scriptscriptstyle C}$ \cite{Campbell_1969}. It is analogous to the London penetration depth, $\lambda_{\rm\scriptscriptstyle L}$, but in the presence of vortices produced by a DC magnetic field. The total measurable response in the mixed state is given by $\lambda_{\rm\scriptscriptstyle m}^2=\lambda_{\rm\scriptscriptstyle L}^2+\lambda_{\rm\scriptscriptstyle C}^2$, where  $\lambda_{\rm\scriptscriptstyle L}$ is the London penetration depth \cite{CoffeyClem,Brandt_1995}. 
Originally, Campbell theory was formulated as a phenomenological description of the magnetic linear AC response in the presence of vortex pinning. Only recently has a general quantitative theory of the Campbell response been developed \cite{Willa2015b, Willa2016} on the basis of the strong pinning theory \cite{Labusch_1969,Larkin_1979,Koopmann2004}. This agrees well with experimental studies \cite{Prozorov_2003_PRB_Campbell_BiSrCuo,Willa2015a}. 

Measurement of the true Campbell length is not a simple task. The sensitivity of most AC and DC susceptometers is insufficient to detect a signal in the Campbell regime. Some of the first reliable measurements of $\lambda_{\rm\scriptscriptstyle C}(T,H,t)$ were performed using a sensitive frequency-domain tunnel diode resonator \cite{Prozorov2000PRB,Giannetta2021} with an AC excitation field of less than \unit[20]{mOe} operating at a frequency around \unit[10-20]{MHz} and with an exceptional resolution of the penetration depth on the single angstrom scale for typical mm-sized samples \cite{Prozorov_2003_PRB_Campbell_BiSrCuo,Propmann_PRB_2011,Willa2015a,hynshoo_2021}. This technique was developed to probe the structure of the superconducting order parameter by measuring $\lambda_{\rm\scriptscriptstyle L}(T)$ \cite{Prozorov2006SST,prozorov2011RPP_review,Carrington2011}, and it also proved to be well suited for Campbell length measurements \cite{Prozorov_2003_PRB_Campbell_BiSrCuo,Willa2015a}.

More recently, new theoretical studies have analyzed the effects of thermal fluctuations (through vortex creep) on the Campbell response of type-II superconductors \cite{Gaggioli2021}. It was realized that, because of the bistability of the free and trapped configurations, thermal fluctuations do assist vortices in escaping from the pinning centers, but also bring vortices into deeper potential wells, essentially extending the reach of the defects and effectively enhancing the average curvature of the pinning potential that determines the Campbell length. This leads to an initial decrease of $\lambda{\scriptscriptstyle C}(T,H,t)$ and, only later, to an increase of the Campbell length when the contribution of the escape events become dominant. In the same picture, the enhanced trapping of vortices has a weaker effect on the average pinning force \cite{Gaggioli2021}, leading to the monotonic decrease of persistent current and DC magnetization observed in experiments. 

This quite general and fundamental theoretical picture of flux creep in the presence of a random pinning potential was never probed or proved experimentally. In this paper, we do so by investigating the time evolution of the Campbell length $\lambda_{\rm \scriptscriptstyle C}$ in the zero-field-cooled protocol with measurements taken on warming (ZFCW) and in the field-cooled protocol (FC). We observed nonmonotonic behavior only within the ZFCW state, exactly as predicted by the strong pinning theory \cite{Gaggioli2021}. 

To investigate whether the general picture of vortex trapping and escaping induced by flux creep is valid, we selected a known isotropic (cubic) type-II superconductor Ca$_3$Ir$_4$Sn$_{13}$, with superconducting transition, $T_c \approx \unit[7.2]{K}$, upper critical field $H_{c2}(0)=\unit[8]{T}$ and London penetration depth, $\lambda_L(0) = \unit[351]{nm}$ \cite{biswas2014superconducting,Santosh_IOP_Ca3Ir4Sn13_2018,wang2015,KrenkelPRB2022}.

\section{Experimental\label{Experimental}}

Single crystals of Ca$_3$Ir$_{4}$Sn$_{13}$ were grown using a high-temperature self-flux method and characterized by X-ray diffraction and energy-dispersive spectroscopy \cite{WangCedomir2012}. This material is fully gapped as concluded from numerous studies, including thermal conductivity \cite{Zhou2012}, specific heat \cite{WangCedomir2012}, and our own London penetration depth measurements \cite{KrenkelPRB2022}. Therefore, the low-temperature London penetration depth is exponentially attenuated. Any temperature variations detected in the vortex state are due to a persistent current density that includes flux creep.

The temperature-dependent variation of the magnetic penetration depth, $\Delta \lambda_{\rm\scriptscriptstyle m}$, was measured using a sensitive self-oscillating tunnel-diode resonator (TDR) technique \cite{VanDegrift1975RSI,Giannetta2021}. The technical details of the measurement apparatus and the calibration are provided elsewhere \cite{Prozorov2000PRB,Prozorov2000APL,Prozorov2021}. An overview of some results from this technique to the study of different superconductors can be found in the review articles \cite{Prozorov2006SST,prozorov2011RPP_review,Carrington2011}. The application of the TDR technique to study the Campbell length can be found in dedicated papers \cite{Prozorov_2003_PRB_Campbell_BiSrCuo,Willa2015a}.

Briefly, the TDR setup consists of a self-oscillating circuit resonating at approximately 14 MHz and producing an AC excitation field of less than \unit[20]{mOe}. When a sample is inserted into the inductor, the total inductance changes resulting in the frequency shift from the empty resonator frequency, $f_0$, proportional to the magnetic susceptibility of the sample, $(f-f_0)/f_0=-G\chi$ where $G$ is the calibration constant, and $4 \pi \chi=\lambda_{\rm\scriptscriptstyle m}/R\tanh{R/\lambda_{\rm\scriptscriptstyle m}}-1$. The effective sample dimension, $R$, is determined by the aspect ratio and sample size \cite{Prozorov2021}. For example, for an infinite cylinder in parallel magnetic field, $R$ coincides with the radius of the cylinder. Our particular sample had dimensions, $565\times575\times255\:\mu m$ and its effective is $R=104\:\mu m$, determined from the established calibration procedure \cite{Prozorov2021}. Importantly, with these dimensions, $\tanh{R/\lambda_{\rm\scriptscriptstyle m}}\approx1$ practically in the whole temperature range, up to roughly  $0.95T_c(H)$, so for all practical purposes we can use the linearized equation, $4 \pi \chi=\lambda_{\rm\scriptscriptstyle m}/R-1$. The sample temperature was actively stabilized within $\pm 10~\text{mK}$ while performing relaxation measurements.

Without an applied external DC magnetic field, in a pure Meissner state, $\lambda_{\rm\scriptscriptstyle m}=\lambda_{\rm\scriptscriptstyle L}$. Applying a DC magnetic field using an external superconducting magnet induces a mixed vortex state, and the measured penetration depth includes both the Campbell and London ones. The Campbell length is evaluated from $\lambda_{\rm\scriptscriptstyle C}^2=\lambda_{\rm\scriptscriptstyle m}^2-\lambda_{\rm\scriptscriptstyle L}^2$.
A small AC excitation field ensures the linear AC response of the vortex lattice, that is, that the vortex oscillations remain well within the trapping range of the pining potential. According to Campbell's phenomenological theory \cite{Campbell_1969, Campbell_1971}, $\lambda_{\rm \scriptscriptstyle C} \propto 1/\sqrt{\alpha}$ where $\alpha$ is the Labusch parameter defined as the curvature of the assumed parabolic pinning potential $U(r)=-\alpha r^2/2$. Within microscopic strong pinning theory, $\alpha$ is shown to be the average pinning curvature experienced by vortices in the random defect landscape \cite{Willa2015a, Willa2015b, Willa2016}. Therefore, the measurements of the Campbell penetration depth provide direct insight into the pinning potential and the mechanisms by which vortices interact with it, which is difficult to access otherwise. This experimental information is crucially important for models of vortex pinning and vortex dynamics.

Since  Ca$_3$Ir$_{4}$Sn$_{13}$ is a relatively clean superconductor, and to examine the effect of different levels of pinning, non-magnetic point-like defects were induced in a controlled way by irradiation of the same sample with 2.5 MeV electrons \cite{KrenkelPRB2022}. The irradiation was performed at the ``SIRIUS" accelerator in the Laboratoire des Solides Irradi\'{e}s at \'{E}cole Polytechnique, Palaiseau, France. Relativistic 2.5 MeV electrons knock out ions, creating vacancy - interstitial Frenkel pairs \cite{Damask1963, Thompson1969}. During irradiation, the sample is kept in liquid hydrogen at around 20 K to prevent immediate recombination and clustering of the atomic defects produced. The acquired irradiation dose is determined by measuring the total charge collected by a Faraday cup located behind the sample. The acquired dose is measured
in ``natural'' units of $\unit[1]{C/cm^2} \equiv 1/e \approx \unit[6.24\times10^{18}]{electrons/cm^2}$. Upon warming the sample to room temperature, some Frenkel pairs recombine and displaced ions diffuse to various sinks (dislocations, surfaces, etc.), leaving a metastable population of point-like defects \cite{PRX, BaKMathesson,KrenkelPRB2022}. The presence of additional scattering is quantitatively verified by measuring the electrical resistivity of the same sample before and after electron irradiation \cite{KrenkelPRB2022}.

\section{Results}\label{results}
We employed two different protocols to measure the time-dependent magnetic relaxation of the Campbell length. The first, often used to study superconductors, is a typical zero-field-cooled protocol with measurements recorded on warming (ZFCW). Here, a sample is cooled in a zero DC magnetic field to the base temperature, around 400 mK in our case. Then a specified DC magnetic field is applied and the measurements are taken while warming, eventually reaching temperatures above $T_c$.  The ZFCW relaxation of the Campbell length is measured by stopping the heating at a target temperature for a certain period of time, up to several hours, during which the data have been continuously collected. Within this protocol, the time-dependent relaxation of $\lambda_{\rm\scriptscriptstyle C}(T,H,t)$ was found to be nonmonotonic for different choices of target temperatures.

The second is a field-cooled (FC) protocol when the measurements are performed on cooling from above $T_c$ without changing the applied DC field. The relaxation of the Campbell length is measured by stopping the temperature sweep at a target temperature and waiting while collecting the data. Field-cooled relaxation of the Campbell length is a  monotonic function of time. The measurements were repeated multiple times, so that only one relaxation pause was made in each run. The curves from multiple runs completely overlap, indicating excellent stability of our set-up and showing that the observed time-dependent effects are due to intrinsic physics. 

\begin{figure}[tbh]
\includegraphics[width=8.6cm]{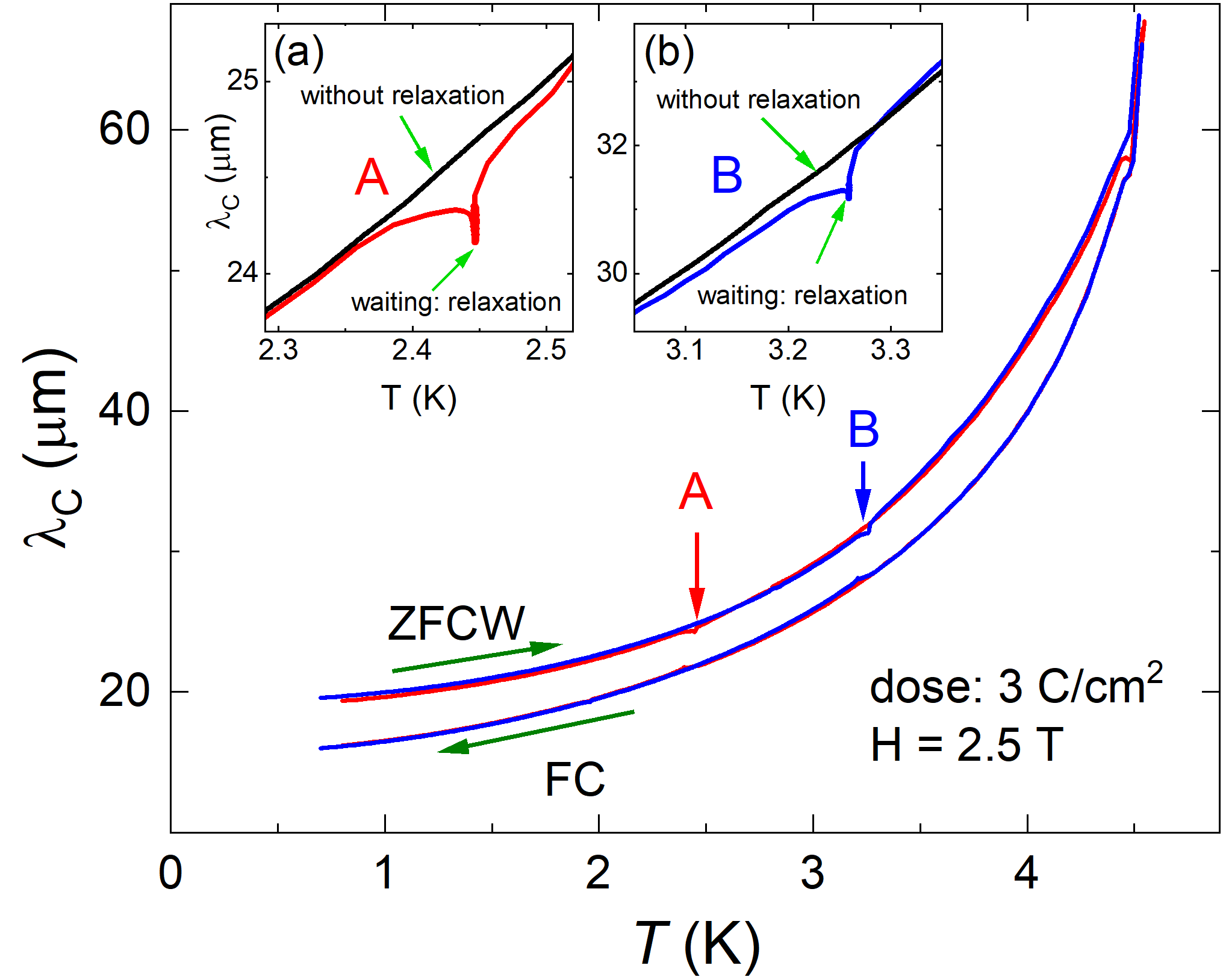} 
\caption{\label{fig1} Temperature dependence of the Campbell length $\lambda_{\rm
\scriptscriptstyle C}$(T) in a sample irradiated with dose of 3 C/cm$^2$. The measurement starts after zero-field cooling to 0.7 K and applying a DC magnetic field of 2.5 T, then taking the data on warming (ZFCW) and then returning back from above $T_c$ in the same field (FC). The relaxation of $\lambda_{\rm \scriptscriptstyle C}(T,H,t)$ was recorded for $\approx$ 40 minutes pausing the ZFCW measurement at 2.45 K (red curve, zoomed in top left inset) and then at 3.25 K (blue curve, zoom in top right inset). Note that these pauses were made during two separate measurements, red and blue curves, respectively. }
\end{figure}

\begin{figure}[tbh]
\includegraphics[width=8.6cm]{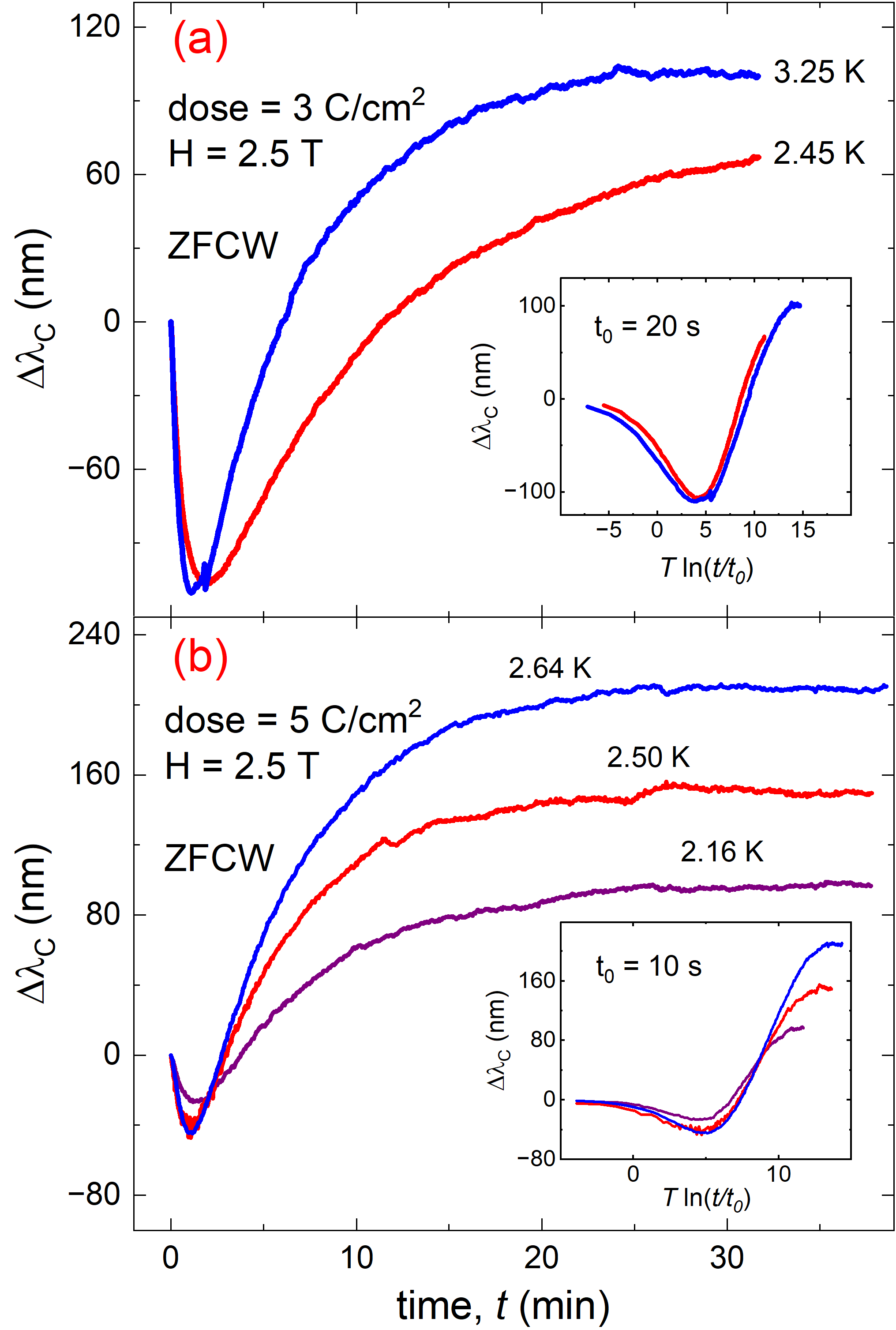} 
\caption{ Main Panel: Nonmonotonic relaxation of $\Delta \lambda_{\rm
\scriptscriptstyle C}(T,H,t)$ at different temperatures along the zero-field-cooled warming (ZFCW) branch. The same sample with different amounts of disorder induced by electron irradiation at (a) 3 C/cm$^2$ and (b) 5 C/cm$^2$. Insets show the same data plotted on a $T\ln{(t/t_0)}$ scale, with the characteristic time, (a) $t_{0} = 20$~s and (b) $t_{0} = 10$~s adjusted to obtain the best scaling.} 
\label{fig2} 
\end{figure}

To understand the contribution of vortex pinning to temperature-dependent relaxation of Campbell length, we started by measuring $\lambda_{\rm\scriptscriptstyle C}(T,H,t)$ in a pristine single crystal of Ca$_3$Ir$_{4}$Sn$_{13}$. In this sample, no significant hysteresis between the zero-field-cooled (ZFCW) and the field-cooled (FC) curves was observed, and magnetic hysteresis $M(H)$ loops were practically reversible. Since a sufficiently large persistent current density is desired to detect relaxation, additional pinning was induced in the same sample by 2.5 MeV electron irradiation with a dose of 3 C/cm$^2$ ($\approx 18.72\times10^{18}$ electrons per cm$^{2}$). After electron irradiation, $T_{c0}$ (at zero magnetic field) was reduced from 7.1 K to 6.95 K, while the upper critical field remained approximately constant, $H_{c2}(0) \approx$ 9 T, consistent with our previous work \cite{KrenkelPRB2022}. Figure~\ref{fig1} shows that, after irradiation, the difference between the ZFCW and FC curves became significant. Two independent runs are shown by red and blue curves. Magnetic relaxation of the Campbell length was measured at several temperatures. Two such relaxations, at $T=2.45\,\textrm{K}$ and $T=3.25\,\textrm{K}$ along the ZFCW branch, are shown in the two insets of Fig.\ref{fig1}. Each vertical curve is a logarithmically slow time dependence of $\lambda_{\rm\scriptscriptstyle C}(T,H,t)$ collected over 40 minutes each. These curves are shown as a function of time in the main panel of Fig.\ref{fig2}. 
In the insets of Fig.\ref{fig1}, the red and blue curves deviate from the unrelaxed curve (black) already before relaxation has taken place and then recover the unrelaxed behavior at some larger temperature.
These observations can be explained as follows:
When approaching the target temperature for relaxation, the heating rate is slowed down in advance to avoid an overshoot. This allows for some part of the relaxation to take place already before the long measurement at constant temperature begins. After relaxation, the Campbell length is initially different from the ZFC value, as the relaxed vortex density gradient implies that $j<j_c(T)$. However, upon further heating, the system returns to the unrelaxed state as soon as $j_c(T)$ falls below $j$ at some temperature $T$ above the target temperature, as predicted by theory \cite{Gaggioli2021}. In other words, this behavior is expected.

\begin{figure}
\includegraphics[width=8.6cm]{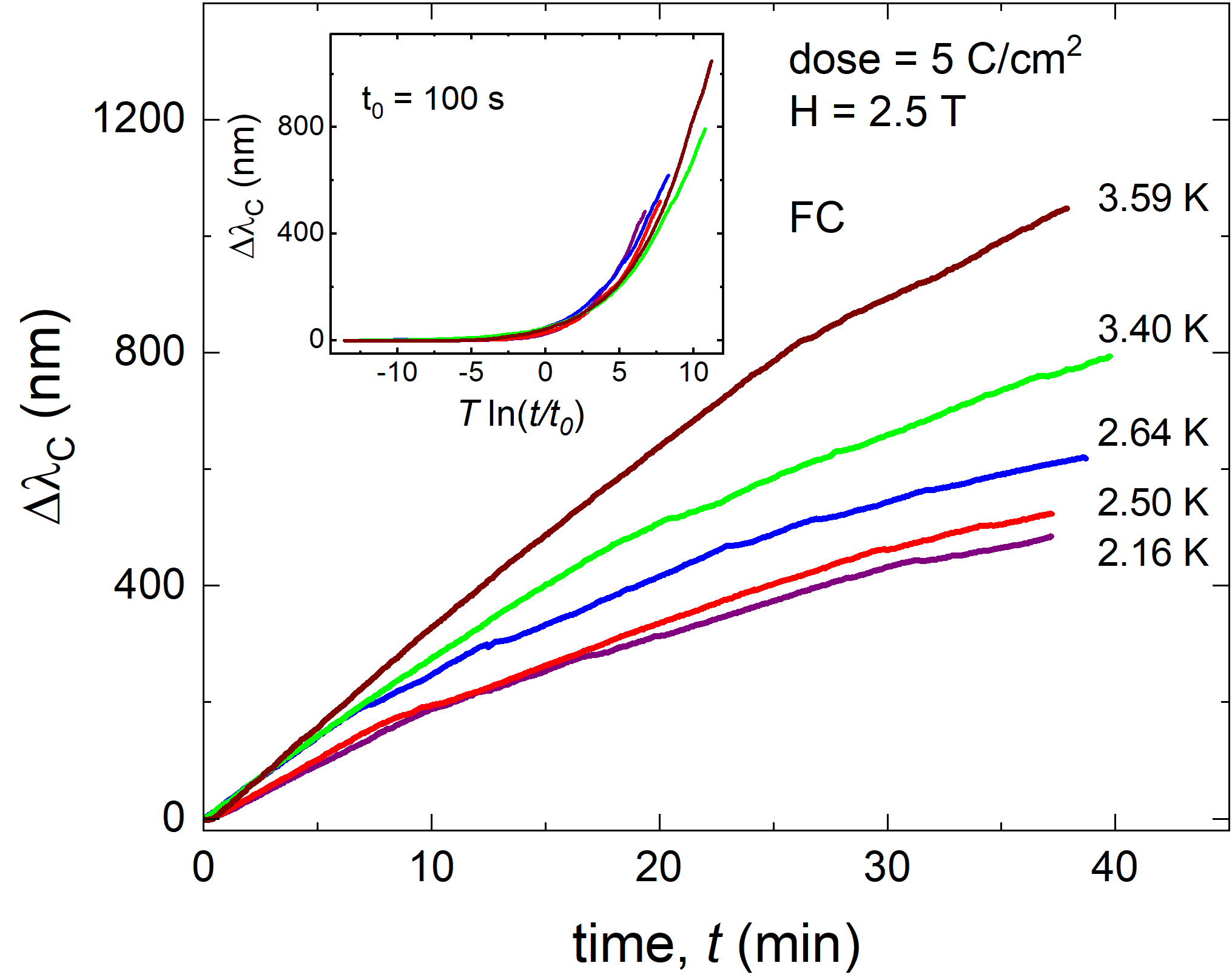} 
\caption{Monotonic time relaxation of $\lambda_{\rm\scriptscriptstyle C}(T,H,t)$ along the field-cooled branch in original time scale. Inset: The characteristic time $t_0 = 100$~s is adjusted to produce the best overlap in logarithmic time scale.}
\label{fig3} 
\end{figure}

Figure~\ref{fig2} zooms in on the change in the Campbell penetration depth during vortex creep, $\Delta \lambda_{\rm
\scriptscriptstyle C}(T,H,t)=\lambda_{\rm
\scriptscriptstyle C}(T,H,t)-\lambda_{\rm
\scriptscriptstyle C}(T,H,0)$ at several temperatures following the ZFCW protocol. The measurements were performed on the same sample with different amounts of pinning induced by electron irradiation. at two doses of (a) and (b) 5 C/cm$^2$.
Figure~\ref{fig2}(a) shows the relaxation curves after the dose of 3 C/cm$^2$ measured at $T=2.45\,\textrm{K}$  (red curve) and $T=3.25\,\textrm{K}$ (blue curve) along the ZFCW branch. 
Note the vertical scale in nanometers, compared to the absolute values of $\lambda_{\rm\scriptscriptstyle C}(T,H,t)$ in micrometers; see Fig.\ref{fig1}. This highlights the smallness of this effect, which is not detectable by conventional DC or AC magnetic measurements. At both temperatures, $\lambda_{\rm \scriptscriptstyle C}(T,H,t)$ shows a non-monotonic relaxation in the ZFCW state. Initially, the Campbell length decreases for more than one minute, and then reverses the behavior, increasing with time. This is not possible in the original Campbell theory, where $\lambda_{\rm \scriptscriptstyle C} \sim 1/\sqrt{j_c}$ is time-independent. To confirm the observed non-monotonic behavior, the same sample was irradiated with an additional dose of 2 C/cm$^2$ to aquire a total dose of 5 C/cm$^2$, and then the magnetic relaxation of the Campbell length was measured at three different temperatures ($2.16\:\text{K}$, $2.50\:\text{K}$, $2.64\:\text{K}$), again on a ZFCW branch, shown in Fig.\ref{fig2}(b). These results confirm the non-monotonic relaxation of Campbell length in the ZFCW branch as predicted by theory \cite{Gaggioli2021}. Figure~\ref{fig3} shows the time-dependent Campbell length at several temperatures measured following a field-cooled (FC) protocol. In stark contrast to ZFCW, the relaxation of Campbell length in an FC branch is monotonic. 

To verify that the observed magnetic relaxation is due to flux creep, the insets of Fig.\ref{fig2} and Fig.\ref{fig3}  show the relaxation curves scaled as a function of the logarithmic time $T\ln{(t/t_0)}$, where $t_0\approx 20\,\text{s}$, $t_0\approx 10\,\text{s}$, and $t_0\approx 100\,\text{s}$ are the respective characteristic relaxation times selected to produce the best overlap of the relaxation curves. This scaling confirms that the relaxation of $\lambda_{\rm \scriptscriptstyle C}(T,H,t)$ is driven by the vortex creep.

\section{Theoretical analysis and discussion}\label{theory}

As introduced by Campbell \cite{Campbell_1969}, the linear response of a type II superconductor to small oscillations of the applied magnetic field $B$ is parameterized by the Campbell length $\lambda_{\rm \scriptscriptstyle C}= \left[B_0^2/4\pi\alpha\right]^{1/2}$ and is determined by the effective curvature of the pinning potential, $\alpha$, acting on the vortex lattice. 
Within strong pinning theory, curvature $\alpha$ is determined by the force jumps $\Delta f_\mathrm{pin}$ experienced by the vortices during trapping and escape events that occur at distances from defects $-x_-^\mathrm{jp}$ and $x_+^\mathrm{jp}$, respectively \cite{Willa2015a,Willa2015b}.
Denoting by $n_p$ and $a_0 \approx \left(\phi_0/B_0\right)^{1/2}$ the volume density of the pinning centers and the inter-vortex distance, the Campbell curvature reads \cite{Gaggioli2021}
\begin{equation}
\label{eq:alpha_finite_T_lk}
   \alpha_\mathrm{sp}(t,T) \approx n_p\frac{\pi}{2}\frac{x^\mathrm{jp}_-}{a_0}
   \frac{\Delta f_\mathrm{pin}(x^\mathrm{jp}_+)
   }{a_0},
\end{equation}
\noindent with the time dependence accounting for the relaxation of $x_\pm^\mathrm{jp}(t, T)$ under the effect of vortex creep.

Equation \eqref{eq:alpha_finite_T_lk} shows the two-fold action of flux creep on the Campbell response. First, the effect of thermal fluctuations is that the vortices have a higher probability of getting trapped. This manifests itself in the growth of the trapping length $x_-^\mathrm{jp}$ with time, leading to an \textit{increase} of $\alpha(t)$ and hence a \textit{decrease} of $\lambda_{\rm \scriptscriptstyle C}(t)$. Second, vortex creep leads to an increase in the number of escape events by weakening the magnitude of the force jump $\Delta f_\mathrm{pin}$ as $x_+^\mathrm{jp}$ is reduced.
In the longer time limit, this leads to a \textit{decrease} of $\alpha$, hence an \textit{increase} of $\lambda_{\rm \scriptscriptstyle C}$, for moderate to large values of the strong pinning parameter $\kappa$. Competition between these opposite effects results in a nonmonotonic behavior of $\lambda_{\rm \scriptscriptstyle C}(T,H,t)$ confirmed by our experiments. 

The application of the strong pinning paradigm can be naturally extended to the FC state. 
In this case, the vortex density gradient is vanishing and the vortex pinning from the defects is symmetric, implying that $x_-^\mathrm{jp} = x_+^\mathrm{jp} = x^\mathrm{jp}$ and no net pinning force is present.
The Campbell curvature $\alpha$, on the other hand, remains finite and is given by \cite{Gaggioli2021}
\begin{equation}\label{eq:alpha_FC}
   \alpha^\mathrm{\scriptscriptstyle FC}_\mathrm{sp} = n_p\frac{\pi\,
   x^\mathrm{jp}}{a_0} \frac{\Delta f_\mathrm{pin}^\mathrm{fp}(x^\mathrm{jp})}{a_0}.
\end{equation}

Since the FC Campbell curvature $\alpha^\mathrm{\scriptscriptstyle FC}_\mathrm{sp}$ in Eq.\eqref{eq:alpha_FC} is a function of a single jump position, the effects of creep on the trapping radius $x^\mathrm{jp}$ and the force jump $\Delta f_\mathrm{pin}(x^\mathrm{jp})$ now work in the same direction and do not lead to a nonmonotonic behavior as observed in the ZFCW state. This agrees well with the monotonic relaxation in the FC state shown in Fig.\ref{fig3}, where $\lambda_{\rm \scriptscriptstyle C}(t)$ only increases when the temperature is kept constant. 

Finally, the excellent agreement between the theory of Ref.\cite{Gaggioli2021} and the present experiments extends beyond the qualitative observation of the monotonic and nonmonotonic time-evolutions characterizing the FC and ZFC states, see main panels of Figs.~\ref{fig2} and \ref{fig3}. A peculiar feature of vortex creep, which distinguishes it from other relaxation mechanisms, is the time-logarithmic behavior due to the current density-dependent barrier for flux jumps. It suggests that relaxation curves measured at different temperatures should collapse under logarithmic rescaling $t\to T\ln t/t_0$, where $t_0$ is some characteristic time scale \cite{Blatter1994}. As shown in inset of Fig.\ref{fig2} and Fig.\ref{fig3}, both the ZFCW and FC experimental relaxation curves overlap quite well in a wide range of temperatures using $t_0\approx 20\,\mathrm{s}$, $t_0\approx 10\,\mathrm{s}$ and $t_0\approx 100\,\mathrm{s}$, respectively. As expected, this occurs regardless of the monotonic or nonmonotonic variation of $\lambda_{\rm \scriptscriptstyle C}$ as this is the general property of the flux creep.

\section{Conclusion} \label{conclusion}

In this work, we experimentally observed a novel and important feature of vortex creep in a random pinning landscape. The probability of vortices jumping out of the potential wells due to thermal fluctuations is higher for shallow wells, but at the same time, thermal diffusion brings more vortices into deeper defects. As a result, in a zero-field-cooled measurement, more and more vortices occupy deeper pinning sites, resulting in a nonmonotonic change of the effective Campbell length, which directly probes the average curvature of the pinning potential. This decrease occurs during macroscopic times of a few minutes in our experiments. The nonmonotonic behavior of the Campbell length is in contrast with the monotonic relaxation of the persistent current and DC magnetization, for which the effect of escaping vortices at $x_+^\mathrm{jp}$ dominates over that of their trapping at $x_-^\mathrm{jp}$. Our observations agree with the predictions of the strong pinning theory.

\section{Acknowledgments}

This work was supported by the National Science Foundation under Grant No. DMR-2219901. M.A.T. and K.R.J. were supported by the U.S. Department of Energy (DOE), Office of Science, Basic Energy Sciences, Materials Science and Engineering Division. Ames National Laboratory is operated for the U.S. DOE by Iowa State University under Contract No. DE-AC02-07CH11358. The authors acknowledge support from the EMIR\&A French network (FR CNRS 3618) on the platform SIRIUS, proposals No. 20-5925 and 23-4663. C.P. acknowledges support by the U.S. Department of Energy, Basic Energy Sciences, Division of Materials Science and Engineering, under Contract No. DE-SC0012704 (BNL). F.G. and V.B.G. acknowledge the financial support of the Swiss National Science Foundation, Division II.
F.G. is grateful for the financial support from the Swiss National Science Foundation (Postdoc.Mobility Grant No. 222230).

%

\end{document}